\documentstyle[12pt]{article}
\begin{document}
\title{\Large\bf Axial and gauge anomalies in the field antifield quantization of the generalized Schwinger model} 

\author{\\
R. Amorim\thanks{\noindent e-mail:
amorim@if.ufrj.br},~~N.R.F.Braga\thanks{\noindent e-mail braga@if.ufrj.br} ~~ and R. Thibes\thanks{\noindent
e-mail: thibes@if.ufrj.br}\\ 
Instituto de F\'{\i}sica\\ 
Universidade Federal do Rio de Janeiro\\
\\
\\}
\date{}

\maketitle
\abstract
In the generalized Schwinger model the vector and axial vector  currents are linearly coupled, with arbitrary coefficients, to the gauge connection. Therefore it represents an interesting example of a theory where both gauge anomalies and  anomalous diver
gences of global currents show up in general. We derive  results for these two kinds of quantum corrections inside the field antifield framework.
\vfill
\noindent PACS: 02.40.-k, 11.10.Ef, 11.15.-q, 11.30.Pb

\vspace{1cm}
\newpage

An anomaly corresponds to the violation, at the quantum level, of some classical symmetry.
In a path integral quantization scheme, this is reflected in the non trivial behavior  of the path integral measure with respect to the corresponding transformation.
It is important to distinguish between anomalies in gauge (local) and in global symmetries. 
Gauge invariance has the important consequence of implying a set of relations among the Greens functions of a field theory, the so called Ward identities, that play a crucial role, for example, in the study of renormalizability.
A quantum obstruction to this kind of symmetry corresponds thus to a failure in 
the process of quantization of a classically gauge invariant theory.
\bigskip

Concerning the case of global symmetries, the Noether theorem implies, at the classical level, that to each global invariance corresponds an associated conserved current. 
Quantum effects can change this picture in such a way that the expectation value of the divergence of a classical conserved current can assume a non vanishing value. Even some non-vanishing divergencies, that so do not correspond to true symmetries of the
 classical action, can have their expectation values modified by quantum effects. This can also be considered as a global anomaly, as it happens with the covariant divergence of the non Abelian axial current in $QCD_4$.
It is important to note that these global anomalous behaviors show up also in field theories that are normally referred in the literature as "non anomalous" (in the sense that they do not have  gauge anomalies). In $QCD_4$,
for instance, adopting a regularization that considers the vector gauge symmetry as a preferred one,
the gauge invariance is not broken at quantum level but the expectation value of the axial current gets anomalous contributions\cite{ABJ,Ja}.
\bigskip

The Lagrangian quantization of Batalin and Vilkovisky (BV) \cite{BV,H,GPS}, also known as field-antifield formalism, is considered to be one of the most powerful procedures of quantization of gauge theories. 
An important feature  of this approach is that, for the case of reducible gauge theories, it furnishes a systematic way of building up the non trivial ghost for
ghost structure.  Another important advantage is that quantum corrections to the path integral measure can be calculated as long as a regularization procedure is introduced.
In this framework, gauge anomalies show up as violations in the so called quantum master equation . The successful application of the Pauli Villars regularization scheme  to this formalism at one loop level\cite{TPN} leads to a series of results on the ca
lculation of gauge anomalies as well as on the implementation of the Wess Zumino mechanism of restoring gauge invariance in the field antifield formalism\cite{GPS,BM1,BM2,GP}.
\bigskip

The calculation of anomalous violations in the conservation of global currents in the field antifield framework was discussed recently in \cite{AB,ABH}.
Those articles consider models where there is no gauge anomaly but just anomalous divergencies of some global currents. There the field antifield formalism is extended by trivially gauging the global symmetries by means of the introduction of  compensatin
g fields in such a way that global anomalies can naturally arise from the generating functional. Different aspects of rigid symmetries inside the field antifield formalism have been discussed in \cite{BHW,HSken}.
We will consider here a specific model where there are in general quantum corrections to both gauge and global symmetries. This  model describes a modified $2D$ electrodynamics where arbitrary linear axial and vector current couplings were introduced and 
it    generalizes that one introduced in \cite{BGZ} . Depending on the parameters chosen, the master equation at one loop order  may have no local solution in the original space of fields and antifields, which then has to be properly extended, following t
he ideas introduced in \cite{FS}.  An interesting feature of our approach is that we can show that it is possible to calculate the anomalous divergencies by means of a canonical transformation and not by introducing compensating fields in order to extend 
the symmetry content of the original classical theory \cite{AB,ABH}. This procedure permits to extract global anomalies even for
models where the gauge symmetries have been restored with the aid of
Wess-Zumino fields.
\bigskip
 
Let us consider the general two dimensional Abelian gauge model,  depending on two arbitrary parameters. This is more general than  the theory of \cite{BGZ} in the sense that it includes also the possibility of pure axial coupling. Its action is given by 
                                        
\footnote{We are using: $\eta_{\mu\nu}={\rm diag.}\,(+1,-1)$;
$\epsilon^{01}=+1$;~ $\{\gamma^\mu,\gamma^\nu\}=2\eta^{\mu\nu}$;~
$\gamma_5=\gamma^0\gamma^1$;~
$\gamma^\mu=\gamma^0\gamma^{\mu\dagger}\gamma^0$;~
$\gamma^\mu\gamma^\nu=\eta^{\mu\nu}+\epsilon^{\mu\nu}\gamma_5$ and
$\gamma_5\gamma^\mu=\epsilon^{\mu\nu}\gamma_\nu$.}

\begin{equation}
\label{model}
{\cal S}_{0}\left[A,\psi,\bar\psi\right]=\int d^2x\,\Big[-\,\frac{1}{4}\,F_{\mu\nu}F^{\mu\nu}
+i\,\bar\psi\, \tilde{ \slash\!\!\!\!D } \psi\Big]\,
\label{e10}
,
\end{equation}
where
\bigskip\noindent

\begin{eqnarray}
F_{\mu\nu}&=&\partial_\mu A_\nu-\partial_\nu A_\mu\,,
\label{e20}\\
\tilde{\slash\!\!\!\!D} &=&\gamma^\mu\,
\big(\partial_\mu-ie\left(s+r\gamma_5\right)A_\mu\big)\,.
\label{e30}
\end{eqnarray}
 
\bigskip\noindent
Action (\ref{model}) reduces to the Schwinger model  when $s=1$, $r=0$; to $2D$ axial electrodynamics when $s=0$, $r=1$ and to the
chiral Schwinger model when $s=1/2$, $r=\pm 1/2$,
but we pose no initial restrictions on the range of variation of these 
parameters. In this sense it describes any interpolation between vector and axial 
$2D$ electrodynamics, displaying in this way an interesting quantum dynamics. The model is classically invariant by the local infinitesimal
transformations

\begin{eqnarray}
\label{local}
\delta\psi&=&ie\alpha(x)\left(s+r\gamma_5\right)\psi\,,\nonumber\\
\delta{\bar\psi}&=&-ie\alpha(x){\bar\psi}\left(s-r\gamma_5\right)\,,\nonumber\\
\delta A_\mu&=&\partial_\mu\alpha(x)\,,
\label{e40}
\end{eqnarray}

\noindent
and also by the global infinitesimal transformations
                               
\begin{eqnarray}
\label{GT}
\delta\psi&=&ie\left(\varepsilon^1+\varepsilon^2\gamma_5\right)\psi\,,\nonumber\\
\delta{\bar\psi}&=&-ie{\bar\psi}\left(\varepsilon^1-\varepsilon^2\gamma_5\right)
\,,
\label{e50}
\end{eqnarray}
where the parameters $\varepsilon^1$ and $\varepsilon^2$ are constant. It is trivial to verify that the transformations (\ref{e40}) close in an Abelian algebra.
The Noether currents associated to the  global transformations (\ref{GT}), with parameters $\varepsilon^1$ and $\varepsilon^2$, are respectively given by

\begin{eqnarray}
\label{Currents}
j^\mu_V&=&
\bar\psi\gamma^\mu\psi\,,
\nonumber\\
j^\mu_A&=&\bar\psi\gamma_5\gamma^\mu\psi\,,
\label{e60}
\end{eqnarray}

\noindent and at classical level the Noether theorem  asserts that
$\,\partial_\mu \,j^\mu_V\,=\,
0\,=\,\,\partial_\mu\, j^\mu_A\,\,$.
\bigskip

Now let us quantize this theory. In the field antifield formalism the generating functional has the general form  \footnote{We are using de Witt's condensed notation which subtends an integral over space time variables when pertinent. Explicitly $\eta_A\P
hi^A\equiv\int d^Dx \eta_A(x)\Phi^A(x)$.}

\begin{equation}
\label{BVF}
Z_\Psi[\eta]=\int [d\Phi^A] \exp\frac{i}{\hbar}
\left\{W\left[\Phi^A,\Phi^*_A=\frac{\partial\Psi}{\partial\Phi^A}\right]+\,\eta_A\Phi^A\,
\right\}\label{1}
\,,
\end{equation}

\noindent where the quantum action is expanded in orders of $\hbar $ as $\,W \,=\,{\cal S}\,+\,\sum_{n\ge1}\hbar^p M_p\,$
and must be a proper solution of the quantum master equation \cite{BV,H,GPS}

\begin{equation}
{1\over2} \left(W,W\right)-i\hbar \Delta W =0
\,.
\label{qme}
\end{equation}

This equation is formally equivalent to the independence of  ({\ref{1}) with respect to the gauge fixing fermion $\Psi$.  
In (\ref{1}) $\Phi^A\,\,$ is a set that includes  the classical fields , ghosts, antighosts, auxiliary fields, etc and   $\Phi^*_A $ are the corresponding antifields. The antibracket $\left(\,\,\,,\,\,\,\right)$ and the  $  \Delta $ operator appearing in 
(\ref{qme}) are defined through

\begin{equation}
\left(X,Y\right)=\frac{\partial^RX}{\partial\Phi^A}\frac{\partial^LY}{\partial\Phi^*_A}-
\frac{\partial^RX}{\partial\Phi^A_*}\frac{\partial^LY}{\partial\Phi_A}
\label{4}
\,,
\end{equation}
\begin{equation}
\Delta X={(-1)}^{\epsilon_A+1}\frac{\partial^R\partial^R}{\partial\Phi^A\partial\Phi^*_A}X
\label{5}
\,,
\end{equation}

\noindent for arbitrary functions $X$ and $Y$ of the fields and antifields.
The corresponding classical action is $ {\cal S}_0\equiv  {\cal S} \left[\Phi^A,\Phi^*_A=0\right]\,$.
\bigskip

Going back to our action (\ref{model}), corresponding to the local symmetry (\ref{e40}) we introduce the ghost $c$ and, as usual, an auxiliary trivial pair $\,{\bar c}\,,\,b\,$ and 
write down the BV action as

\begin{eqnarray}
\label{MBV}
{\cal S}&=&{\cal S}_0+\int d^2 x\,\Big[{A^*}^\mu\partial_\mu c
-ie\psi^*\left[\left(s+r\gamma_5\right)c
\right]\psi\nonumber\\
&&+ie\bar\psi\left[\left(s-r\gamma_5\right)c
\right]\bar\psi^*+{\bar c}^*b
\Big]\,.
\label{e70}
\end{eqnarray}

In order to solve the master equation (\ref{qme}) at one loop level  we must calculate $ \Delta {\cal S}$. In the field antifield formalism, at this loop order, the Pauli Villars (PV) regularization prescription can be easily adopted . We add to the actio
n (\ref{MBV}) a PV action with fields $\bar\chi$ and $\chi$ and choose, as usual, a mass term of the form $ \,\,{\cal M}\bar\chi \chi\,\,$ that considers the vector symmetry as a privileged one. Following then the steps described in ref \cite{GPS,TPN} we 
find

\begin{equation}
\label{Deltas}
\Delta {\cal S}=\frac{ie^2}{\pi}\int d^2x \, rc\left(
r\partial_\mu A^\mu+s\epsilon^{\mu\nu}\partial_\mu A_\nu
\right)
\,.
\label{e80}
\end{equation}

The term $\epsilon^{\mu\nu}\partial_\mu A_\nu$ is cohomologically non trivial while the term $\partial_\mu A^\mu$ is  trivial \cite{H}.
If $r=0$, $ \Delta {\cal S} $ vanishes identically, which is the expected value,
since according Eq. (\ref{e30}), we would have in this case a pure vector current coupling  with the gauge connection and we are adopting a regularization that takes the vector symmetry as a preferred one. If $r\neq0$ but
$s\,=\,0$ there is no genuine anomaly in the theory as we easily see that 

\begin{equation}
M_1\,=\,{e^2\,r^2\over 2 \pi}\,\int d^2x \Big( A_\mu A^\mu \Big).
\end{equation}

\noindent solves the master equation. This result in some sense could be expected   due to the $2D$ identity $\gamma_5\gamma^\mu=\epsilon^{\mu\nu}\gamma_\nu$. It implies that the axial current coupling term corresponds to a vector current coupling with a 
modified "gauge connection" given by $\tilde A_\mu=\epsilon_{\mu\nu}A^\nu$.
The anomaly under an axial gauge transformation (See Eq. (\ref{e40}).) should then
be proportional to $\epsilon_{\mu\nu}\tilde F^{\mu\nu}$, where $\tilde F_{\mu\nu}$ is the "curvature"associated to $\tilde A_\mu$. But 
$\epsilon_{\mu\nu}\tilde F^{\mu\nu}=2\partial_\mu A^\mu$,
which is just the cohomological trivial term found in the present situation. Of course this is not valid for higher dimensions.
\bigskip

In the general case, however, when $r\neq0$ and $s\neq0$, no $M_1$ can be found (in the original space of fields)  which cancels $ \Delta S $ in the first loop order term of the expansion of (\ref{qme}). Thus we are in presence of a true gauge anomalous t
heory. 
If we take the point of view of references \cite{FS} and enlarge the set of fields of the theory by means of the introduction of a Wess Zumino field $\theta$ that transforms exactly in the original gauge group as $ \delta\theta (x) \,=\,\alpha (x)\,$, we 
can write
a solution to the master equation as

\begin{eqnarray}
M_1&=&\frac{ie^2}{\pi}\int d^2x
\left\{
\frac{1}{2}ar^2A_\mu A^\mu+\frac{1}{2}r^2(a-1)\partial_\mu\theta \partial^\mu\theta+
\right.\nonumber\\&&\left.
\theta\left[r^2(a-1)\partial_\mu A^\mu -rs\epsilon^{\mu\nu}\partial_\mu A_\nu\right]\right\}
\label{e90}
\,.
\end{eqnarray}

Since $ \delta M_1 - i\Delta {\cal S}=0$, it follows that (\ref{qme}) is satisfied for
\begin{equation}
W={\cal S}+\hbar M_1+\int d^2\,x\theta^*c
\,.
\label{e100}
\end{equation}

\noindent This kind of procedure was proposed in the field antifield framework in \cite{BM2,GP}.
The arbitrary parameter $a$ which appears in (\ref{e100}) is to be identified with the Jackiw Rajaraman parameter\cite{JR}.
\bigskip
 
Let us now look at the global symmetries. First let us show how can we calculate the anomalous divergencies of generic global currents by canonical transformations. If  we perform an infinitesimal  transformation generated by

\begin{equation}
F(\Phi^A,\Phi^{*'}_A)=\Phi^{*'}_A\Phi^A +f(\Phi^A,\Phi^{*'}_A)
\label{7}
\end{equation}
with
\begin{equation}
f=\epsilon^\alpha\Phi^{*'}_AG^A_\alpha[\Phi]
\,,
\label{f}
\end{equation}

\noindent where the $\epsilon^\alpha$'s are some arbitrary infinitesimal local parameters, the fields $\Phi^A$, $\Phi^*_A$ transform as 

\begin{eqnarray}
\Phi^{A'}=&\frac{\partial^LF}{\partial \Phi^{*'}_A}&=\Phi^A+\epsilon^\alpha G^A_\alpha[\Phi]\,,
\label{9}\\
\Phi^*_A=\frac{\partial^R F}{\partial\Phi^A}&\rightarrow&
\Phi_A^{*'}=\Phi^*_B\left(\delta^B_A-\epsilon^\alpha\frac{\partial^R G_\alpha^B}{\partial\Phi^A}\right)
\label{10}
\,.
\end{eqnarray}

It is possible to show \cite{GPS,TPN} that the new action  

\begin{equation}
{\tilde W}=W' -i\hbar \Delta f\,,
\label{11}
\end{equation}

\noindent where $W'$ is the action $W$ written in terms of the new primed variables, also satisfies (\ref{qme}).

The factor $ \Delta f $ compensates the possible noninvariance of the measure of integration of (\ref{1}) with respect to the canonical transformation. However, exactly as it happens in the case of $ \Delta S $, the action of the operator $ \Delta $ must 
be regularized.
When the regularized result is non vanishing, it can be put in the form

\begin{equation}
-i\Delta f= \epsilon^\alpha{\cal A}_\alpha [\phi^i]
\,,
\label{a}
\end{equation}

A generating functional $Z_\Psi[J,\epsilon]$ constructed with ${\tilde W}$ naturally leads to the same quantized theory and since the parameters $\epsilon^\alpha$ are  arbitrary, we must have the identities

\begin{equation}
\frac{\partial Z_\Psi[J,\epsilon]}{\partial\epsilon^\alpha(x)}\equiv 0
\label{12}
\,.
\end{equation}

Whenever $ \Delta f \neq 0 $ these identities may be used to derive anomalous results involving ${\cal A}_\alpha$. Introducing:

\begin{equation}
j^\mu_\alpha\,=\,{\partial W^\prime\over \partial ( \partial_\mu \epsilon^\alpha)}\,,
\end{equation}

\noindent we get the general relation involving the quantum expectation value of the divergence of this current

\begin{equation}
\frac{\partial Z_\Psi[J,\epsilon]}{\partial\epsilon^\alpha(x)}\,=\,
{i\over \hbar}\,\langle\,-\,\partial_\mu j^\mu_\alpha\,+\,
{\partial W^\prime\over \partial \epsilon^\alpha} \,-\,\hbar {\cal A}_\alpha\,
\,-\,G^A_\alpha [\Phi] \eta_A \, \rangle
\end{equation}

Therefore, to calculate the anomalous contributions to the divergence of our global currents (\ref{Currents}) we perform a canonical transformation generated by (\ref{7}) with

\begin{equation}
f=\,- i\epsilon^1\left(\psi^{\ast'}\psi\,-\, {\bar\psi}{\bar\psi}^{\ast'}\right)\,
-\,i\epsilon^2\left(\psi^{\ast'}\gamma_5 \psi\,+\, 
{\bar\psi}\gamma_5 {\bar\psi}^{\ast'}\right)
\label{e110}\,,
\end{equation}

\noindent that incompasses both vectorial and axial transformations

\begin{eqnarray}
\psi'&=&\left(1\,-\, i \epsilon^1 \,-\,i\epsilon^2 \gamma_5\right)\psi\,,\nonumber\\
{\bar\psi}'&=&{\bar\psi}\left(1\,+\,i\epsilon^1\,-\, i\epsilon^2 \gamma_5\right)\,,\nonumber\\
{\psi^*}{}'&=&\psi^*\left(1\,+\, i \epsilon^1\,+\, i\epsilon^2 \gamma_5\right)\,,\nonumber\\
{{\bar\psi}^*}{}'&=&\left(1\,-\,i \epsilon^1 \,+\, i\epsilon^2\gamma_5\right)
{\bar\psi}^*
\label{e120}
\,,
\end{eqnarray}

\noindent keeping the remaining fields unchanged.

Assuming a regularization analogous to that one adopted for the calculation of (\ref{e80}), 
the derivation of  $ \Delta f$ is analogous to  that one of $ \Delta S$ in  with $rc$  is replaced by $\epsilon$ and it gives
\begin{equation}
\Delta f=\frac{ie^2}{\pi}\int d^2x \, \epsilon\left(
r\partial_\mu A^\mu+s\epsilon^{\mu\nu}\partial_\mu A_\nu
\right)
\,.
\label{e130}
\end{equation}

We will assume that, as usual,  the gauge fixing fermion $\Psi$ in eq. (\ref{BVF}) does not depend on the fermionic fields. This is equivalent to setting  

\begin{equation}
\bar\psi^\ast\,=\,\psi^\ast\,=\,0
\end{equation}

\noindent at the gauge fixed level (after the canonical transformations are performed).

Substitution of (\ref{e120}) and (\ref{e130}) in (\ref{11}) leads through (\ref{12}) to

\begin{eqnarray}
<\partial_\mu j^\mu_A>\vert_{_{\eta=\bar\eta=0}}&=&\frac{\hbar e^2}{\pi}<\int d^2 x \left(r\partial_\mu A^\mu+s\epsilon^{\mu\nu}\partial_\mu A_\nu\right)>
\nonumber\\
<\partial_\mu j^\mu_V>\vert_{_{\eta=\bar\eta=0}}&=& 0
\label{e140}
\,,
\end{eqnarray}
where $\eta$ and $\bar\eta$ are the external sources corresponding to $\psi$ and $\bar\psi$ respectively.

These results, together with equation (\ref{Deltas}) show us that our model has, in general, both gauge and global anomalies (violation in the conservation of global currents).
The fact that the vector current is always conserved clearly reflects the choice of a mass term for the Pauli Villars regulating fields that preserves the vector symmetry. It is interesting to observe that the axial current will always have an anomalous d
ivergence even in the cases particular cases where there is no gauge anomaly. 
\bigskip

It is interesting to see what happens in the non Abelian case. The non Abelian version of (\ref{e10}) reads

\begin{equation}
\label{NAA}
{\cal S}_{0}\left[A,\psi,\bar\psi\right]=\int d^2x\,\Big[-\,\frac{1}{4}\mbox{ tr } F^{\mu\nu} F_{\mu\nu}\,+\,i\,\bar\psi\,\slash\!\!\!\!{\tilde D}\psi\Big]\,
\label{e150}
\end{equation}

\noindent where now we have

\begin{eqnarray}
\slash\!\!\!\!{\tilde D} &=&\gamma^\mu\,
\big(\partial_\mu-ie\left(s+r\gamma_5\right)A_\mu\big)\,,\nonumber\\
F_{\mu\nu}&=&\partial_\mu A_\nu-\partial_\nu A_\mu-ie\left[A_\mu,A_\nu\right]\,,\nonumber\\
A_\mu&=&A_\mu^aT^a\,,\nonumber\\
\left[T^a,T^b\right]&=&if^{abc}T^c\,,\nonumber\\
\mbox{tr }T^aT^b&=&\delta^{ab}\,.
\label{e160}
\end{eqnarray}

The natural generalization of the transformations (\ref{local}) reads

\begin{eqnarray}
\delta\psi&=&ie\left(s+r\gamma_5\right)\alpha\psi\,,\nonumber\\
\delta{\bar\psi}&=&-ie{\bar\psi}\left(s-r\gamma_5\right)\alpha\,,\nonumber\\
\delta A_\mu&=&\partial_\mu\alpha-ie\left[A_\mu,\alpha\right]\,,
\label{e170}
\end{eqnarray}

\noindent where now the infinitesimal parameters $\alpha\equiv\alpha^aT^a$  takes values at the Lie algebra of the group generated by $T^a$.
However, a simple calculation shows that these transformations leave the action (\ref{NAA}) invariant if and only if 

\begin{equation}
{\left(s+r\gamma_5\right)}^2=\left(s+r\gamma_5\right)
\label{e180}
\,.
\end{equation}

\noindent This restricts the range of variations of the pair $(s,r)$ to  $(1,0)$ or $(\frac{1}{2},\pm \frac{1}{2})$ (besides the trivial $(0,0)$). So, in the non Abelian case, action (\ref{NAA}) does
not represent any new general model but only a concise representation for chiral or vector
$QCD_2$. There is an extensive literature about these two models \cite{QCD2}, to what we
refer. It would be possible, however, to apply the methods here described, as well as those found in \cite{AB,ABH}, to any of these two non Abelian models. 
\vskip 1cm
\noindent {\bf Acknowledgment:} This work is partially supported  by
 CNPq, FINEP and FUJB (Brazilian Research Agencies).


\begin{thebibliography}{30}
\bibitem{ABJ}S. Adler , Phys. Rev. 177 (1969) 2426;
J. Bell and R. Jackiw, Nuovo Cim. 60A (1969)  47.
\bibitem{Ja} For a review, see R. Jackiw, in Lectures on Current Algebra and it's Applications", ed. S. Treiman et al.
, Princeton University Press, Princeton, NJ, 1972.
\bibitem{BV} I.A. Batalin and G.A. Vilkovisky, Phys. Lett. B102
(1981) 27; Phys. Rev. D28 (1983) 2567.
\bibitem{H} For a review, see M.Henneaux and C. Teitelboim,
{\it Quantization of gauge systems} (Princeton University Press, New
Jersey, 1992).
\bibitem{GPS} A recent review with a wide list of references
can be found in J. Gomis, J. Paris and S. Samuel, Phys. Rep. 259
(1995) 1.
\bibitem{TPN}  W.Troost, P.van Nieuwenhuizen and
A. Van Proeyen, Nucl. Phys. B333 (1990) 727.
\bibitem{BM1} N. F. R. Braga and H. Montani, Phys. Lett. B264 (1991) 125
\bibitem{BM2} N. F. R. Braga and H. Montani,  Int. J. Mod. Phys. A8 (1993) 2569.
\bibitem{GP} J. Gomis, J. Paris Nucl. Phys. B395 (1993) 288.
\bibitem{AB} R. Amorim and N. R. F. Braga, Phys. Review D57 (1998)1225.
\bibitem{ABH}R. Amorim , N. R. F. Braga and M. Henneaux , Phys. Lett. B436 (1998) 125.
\bibitem{BHW} F. Brandt, M. Henneaux and A. Wilch,
Nucl.  Phys.  B 510 (1998) 640.
\bibitem{HSken} T. Hurth and K. Skenderis, "Quantum Noether Method",
hep-th/9803030.
\bibitem{BGZ} A. Bassetto, L. Griguolo and P. Zanca, Phys. Rev. D 50
(1994) 1077.
\bibitem{FS} L. D. Faddeev, Phys. Lett. B145
(1984) 81; L. D. Faddeev and S. L. Shatashvili, Phys. Lett. B167
(1986) 225.
\bibitem{JR} R. Jackiw and R. Rajaraman, Phys. Rev. Lett. 54 (1985) 1219.
\bibitem{QCD2} See for instance E. Abdalla, M. C. B. Abdalla, Phys. Rept. 265 (1996) 253 and References therein.
\end{thebibliography}
\end{document}